\documentclass{svproc}
\usepackage{url}
\usepackage{graphicx}
\usepackage{ccicons}                              

\usepackage{multirow}
\usepackage{amsmath}
\usepackage[colorlinks=true,
            linkcolor=red,
            urlcolor=blue,
            citecolor=blue]{hyperref}
\usepackage{comment}

\begin{document}
\mainmatter              
\title{Diagnosis of Celiac Disease and Environmental Enteropathy on Biopsy Images Using Color Balancing on Convolutional Neural Networks}
\titlerunning{Diagnosis of Celiac Disease and Environmental Enteropathy}
%
%
\author{Kamran Kowsari\inst{1} \and Rasoul Sali\inst{1} \and Marium N. Khan \inst{3} \and William Adorno\inst{1} \and S. Asad Ali\inst{4} \and Sean R. Moore \inst{3} \and Beatrice C. Amadi \inst{5} \and Paul Kelly \inst{5,6}  \and Sana Syed \inst{2,3,4,*} \and Donald E. Brown\inst{1,2,*}}
\authorrunning{Kamran Kowsari et al.} 
%
%
\institute{Department  of  Systems \&  Information  Engineering,  University  of Virginia, Charlottesville, VA, USA
\and
School  of  Data  Science,  University  of  Virginia,  Charlottesville,  VA, USA
\and 
Department of Pediatrics, School of Medicine,  University  of  Virginia,  Charlottesville,  VA, USA
\and
Aga  Khan  University, Karachi, Pakistan
\and
Tropical  Gastroenterology  and  Nutrition  group,  University  of  Zambia School of Medicine, Lusaka, Zambia
\and
Blizard  Institute,  Barts  and  The  London  School  of  Medicine,  Queen Mary University of London, London, United Kingdom\\~\\
$^*$ Co-corresponding authors:~\{\href{mailto:sana.syed@virginia.edu}{sana.syed}, \href{mailto:deb@virginia.edu}{deb}\}@virginia.edu}
\maketitle              
\setcounter{footnote}{0}
\begin{abstract}
Celiac Disease~(CD) and Environmental Enteropathy~(EE) are common causes of malnutrition and adversely impact normal childhood development. CD is an autoimmune disorder that is prevalent worldwide and is caused by an increased sensitivity to gluten. Gluten exposure destructs the small intestinal epithelial barrier, resulting in nutrient mal-absorption and childhood under-nutrition. EE also results in barrier dysfunction but is thought to be caused by an increased vulnerability to infections. EE has been implicated as the predominant cause of under-nutrition, oral vaccine failure, and impaired cognitive development in low-and-middle-income countries. Both conditions require a tissue biopsy for diagnosis, and a major challenge of interpreting clinical biopsy images to differentiate between these gastrointestinal diseases is striking histopathologic overlap between them. In the current
study, we propose a convolutional neural network~(CNN) to classify duodenal biopsy images from subjects with CD, EE, and healthy controls. We evaluated the performance of our proposed model using a large cohort containing 1000 biopsy images. Our evaluations show that the proposed model achieves an area under ROC of 0.99, 1.00, and 0.97 for CD, EE, and healthy controls, respectively. These results demonstrate the discriminative power of the
proposed model in duodenal biopsies classification.
\keywords{Convolutional Neural Networks, Medical Imaging, Celiac Disease, Environmental Enteropathy }
\end{abstract}
\section{Introduction and Related Works}\label{sec:Introduction} 

Under-nutrition is the underlying cause of approximately~$45$\% of the~$5$ million under~$5$-year-old childhood deaths annually in low and middle-income countries~(LMICs)~\cite{WHO.Children} and is a major cause of mortality in this population. Linear growth failure (or stunting) is a major complication of under-nutrition, and is associated with irreversible
physical and cognitive deficits, with profound developmental implications~\cite{syed2016environmental}. A common cause of stunting in
LMICs is EE, for which there are no universally accepted, clear diagnostic algorithms or non-invasive
biomarkers for accurate diagnosis~\cite{syed2016environmental}, making this a critical priority~\cite{naylor2015environmental}. EE has been described to be caused by 
chronic exposure to enteropathogens which results in a vicious cycle of constant
mucosal inflammation, villous blunting, and a damaged epithelium~\cite{syed2016environmental}. These deficiencies contribute to a markedly reduced nutrient absorption and thus under-nutrition and stunting~\cite{syed2016environmental}. Interestingly, CD, a common cause of stunting in
the United States, with an estimated~$1$\% prevalence, is an autoimmune disorder caused by a gluten sensitivity~\cite{husby2012european} and has many shared histological features with EE~(such as increased inflammatory cells and villous blunting)~\cite{syed2016environmental}. This resemblance has led to the major challenge of differentiating clinical biopsy images for these similar but distinct diseases. 
Therefore, there is a major clinical interest towards developing new, innovative methods to automate and enhance the detection of morphological features of
EE versus CD, and to differentiate between diseased and healthy small intestinal tissue~\cite{bejnordi2017diagnostic}. 

\begin{figure}[h]
    \centering
    \includegraphics[width=\columnwidth]{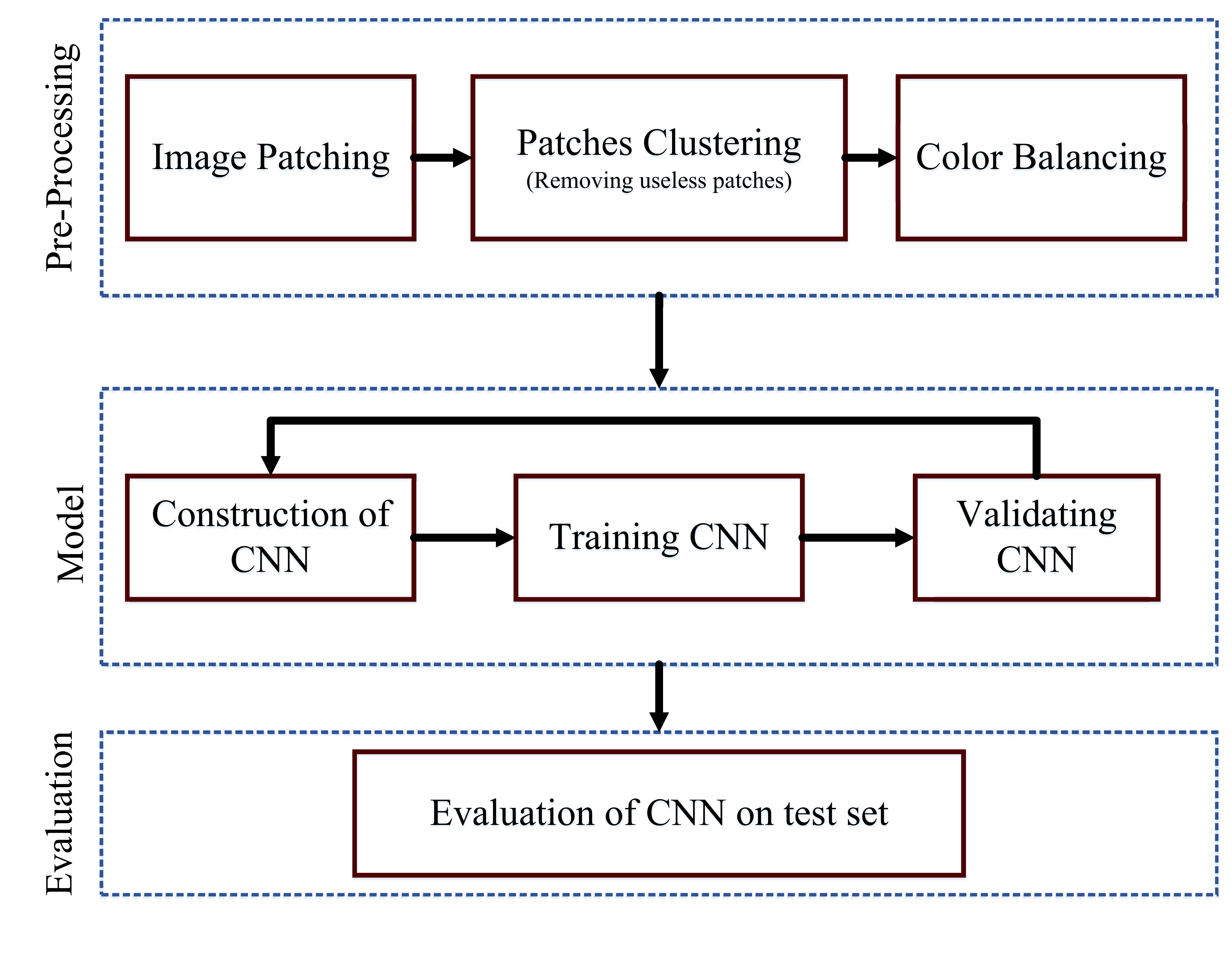}
    \caption{Overview of methodology} \label{fig:Pipeline}
\end{figure}

In this paper, we propose a CNN-based model for classification of biopsy images. In recent years, Deep Learning architectures have received great attention after achieving state-of-the-art results in a wide variety of fundamental tasks such classification~\cite{Heidarysafa2018RMDL,kowsari2017hdltex,kowsari2018rmdl,info10040150,litjens2017survey,nobles2018identification,zhai2016doubly} or other medical domains~\cite{hegde2019comparison,zhang2018patient2vec}. CNNs in particular have proven to be very effective in medical image processing. CNNs preserve local image relations, while reducing dimensionality and for this reason are the most popular machine learning algorithm in image recognition and visual learning tasks~\cite{ker2018deep}. CNNs have been widely used for classification and segmentation in various types of medical applications such as histopathological
images of breast tissues, lung images, MRI images,  medical X-Ray images, etc.~\cite{gulshan2016development,litjens2017survey}. Researchers produced advanced results on duodenal biopsies classification using CNNs~\cite{Mohammad_al_boni}, but those models are only robust to a single type of image stain or color distribution. Many researchers apply a stain normalization technique as part of the image pre-processing stage to both the training and validation datasets~\cite{nawaz2018classification}. In this paper, varying levels of color balancing were applied during image pre-processing in order to account for multiple stain variations. 

The rest of this paper is organized as follows: In Section~\ref{sec:Data_Source}, we describe the different data sets used in this work, as well as, the required pre-processing steps. The architecture of the model is explained in Section~\ref{sec:Method}. Empirical results are elaborated in Section~\ref{sec:Empirical_Results}. Finally, Section~\ref{sec:Conclusion} concludes the paper along with outlining future directions.

\section{Data Source}\label{sec:Data_Source}~ For this project, $121$ Hematoxylin and Eosin (H\&E) stained duodenal biopsy glass slides were retrieved from~$102$ patients. The slides were converted into~$3118$ whole slide images, and labeled as either EE, CD, or normal. The biopsy slides for EE patients were from the Aga Khan University Hospital~(AKUH) in Karachi, Pakistan~($n = 29$ slides from~$10$ patients) and the University of Zambia Medical Center in Lusaka, Zambia ($n = 16$). The slides for CD patients ($n = 34$) and normal ($n = 42$) were retrieved from archives at the University of Virginia~(UVa). The CD and normal slides were converted into whole slide images at~$40$x magnification using the Leica SCN~$400$ slide scanner (Meyer Instruments, Houston, TX) at UVa, and the digitized EE slides were of 20x magnification and shared via the Environmental Enteric Dysfunction Biopsy Investigators~(EEDBI) Consortium shared WUPAX server. Characteristics of our patient population are as follows: the median~($Q1$, $Q3$) age of our entire study population was~$31$~($20.25$, $75.5$) months, and we had a roughly equal distribution of males~($52$\%, $n = 53$) and females~($48$\%, $n = 49$). The majority of our study population were histologically normal controls~$(41.2\%)$, followed by CD patients~$(33.3\%)$, and EE patients~$(25.5\%)$.

\section{Pre-Processing}\label{sec:Pre-Processing}
In this section, we cover all of the pre-processing steps which include image patching, image clustering, and color balancing. The biopsy images are unstructured~(varying image sizes) and too large to process with deep neural networks; thus, requiring that images are split into multiple smaller images. After executing the split, some of the images do not contain much useful information. For instance, some only contain the mostly blank border region of the original image. In the image clustering section, the process to select useful images is described. Finally, color balancing is used to correct for varying color stains which is a common issue in histological image processing.

\subsection{Image Patching}\label{subsec:Clustering}
Although effectiveness of CNNs in image classification has been shown in various studies in different domains, training on high resolution Whole Slide Tissue Images (WSI) is not commonly preferred due to a high computational cost. However, applying CNNs on WSI enables losing a large amount of discriminative information due to extensive downsampling~\cite{hou2016patch}. Due to a cellular level difference between Celiac, Environmental Entropathy and normal cases, a trained classifier on image patches is likely to perform as well as or even better than a trained WSI-level classifier. Many researchers in pathology image analysis have considered classification or feature extraction on image patches~\cite{hou2016patch}.
In this project, after generating patches from each images, labels were applied to each patch according to its associated original image. A CNN was trained to generate predictions on each individual patch.

\begin{figure}[!b]
    \centering
    \includegraphics[width=\textwidth]{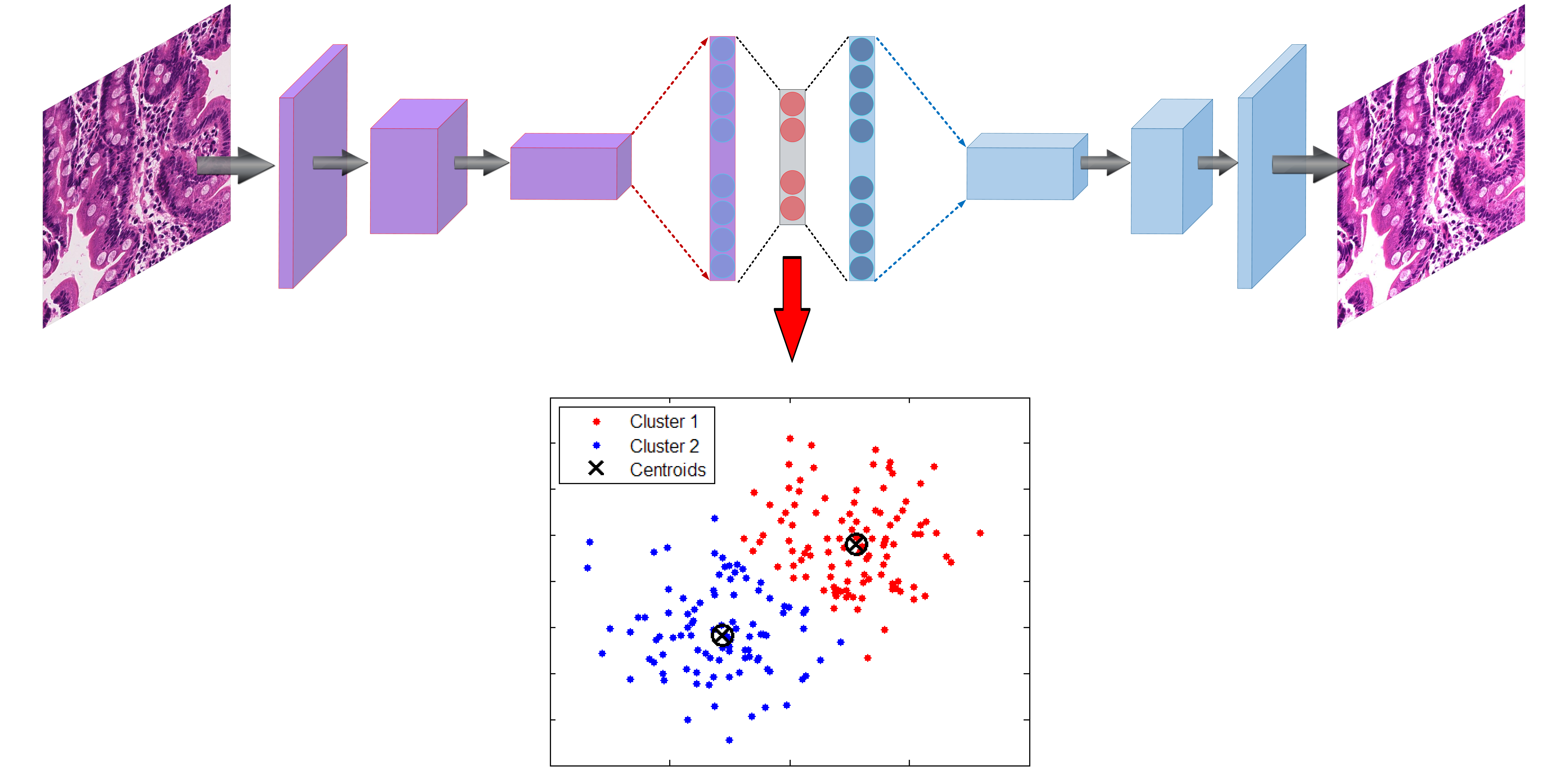}
    \caption{Structure of clustering model with autoencoder and K-means combination} \label{fig:AE}
\end{figure}

\subsection{Clustering}\label{subsec:Clustering}~\\
In this study, after image patching, some of created patches do not contain any useful information regarding biopsies and should be removed from the data. These patches have been created from mostly background parts of WSIs. A two-step clustering process was applied to identify the unimportant patches. For the first step, a convolutional autoencoder was used to learn embedded features of each patch and in the second step we used k-means to cluster embedded features into two clusters: useful and not useful. In Figure~\ref{fig:AE}, the pipeline of our clustering technique is shown which contains both the autoencoder and k-mean clustering.  

An autoencoder is a type of neural network that is designed to match the model's inputs to the outputs~\cite{goodfellow2016deep}. The autoencoder has achieved great success as a dimensionality reduction method via the powerful reprehensibility of neural networks~\cite{wang2014generalized}. The first version of autoencoder was introduced by~\textit{DE. Rumelhart el at.}~\cite{rumelhart1985learning} in 1985. The main idea is that one hidden layer between input and output layers has much fewer units~\cite{liang2017text} and can be used to reduce the dimensions of a feature space. For medical images which typically contain many features, using an autoencoder can help allow for faster, more efficient data processing.

A CNN-based autoencoder can be divided into two main steps~\cite{masci2011stacked}~: encoding and decoding.
\begin{equation}
\begin{split}
    O_m(i, j) = a\bigg(&\sum_{d=1}^{D}\sum_{u=-2k-1}^{2k+1}\sum_{v=-2k -1}^{2k +1}F^{(1)}_{m_d}(u, v)I_d(i -u, j -v)\bigg) \\&\quad m = 1, \cdots, n
\end{split}
\end{equation}
Where~$F \in \{F^{(1)}_{1},F^{(1)}_{2},\hdots,F^{(1)}_{n},\}$ is a convolutional filter, with convolution among an input volume defined by~$I = \left\{I_1,\cdots, I_D\right\}$ which it learns to represent the input by combining non-linear functions:

\begin{figure}[!b]
    \centering
    \includegraphics[width=\columnwidth]{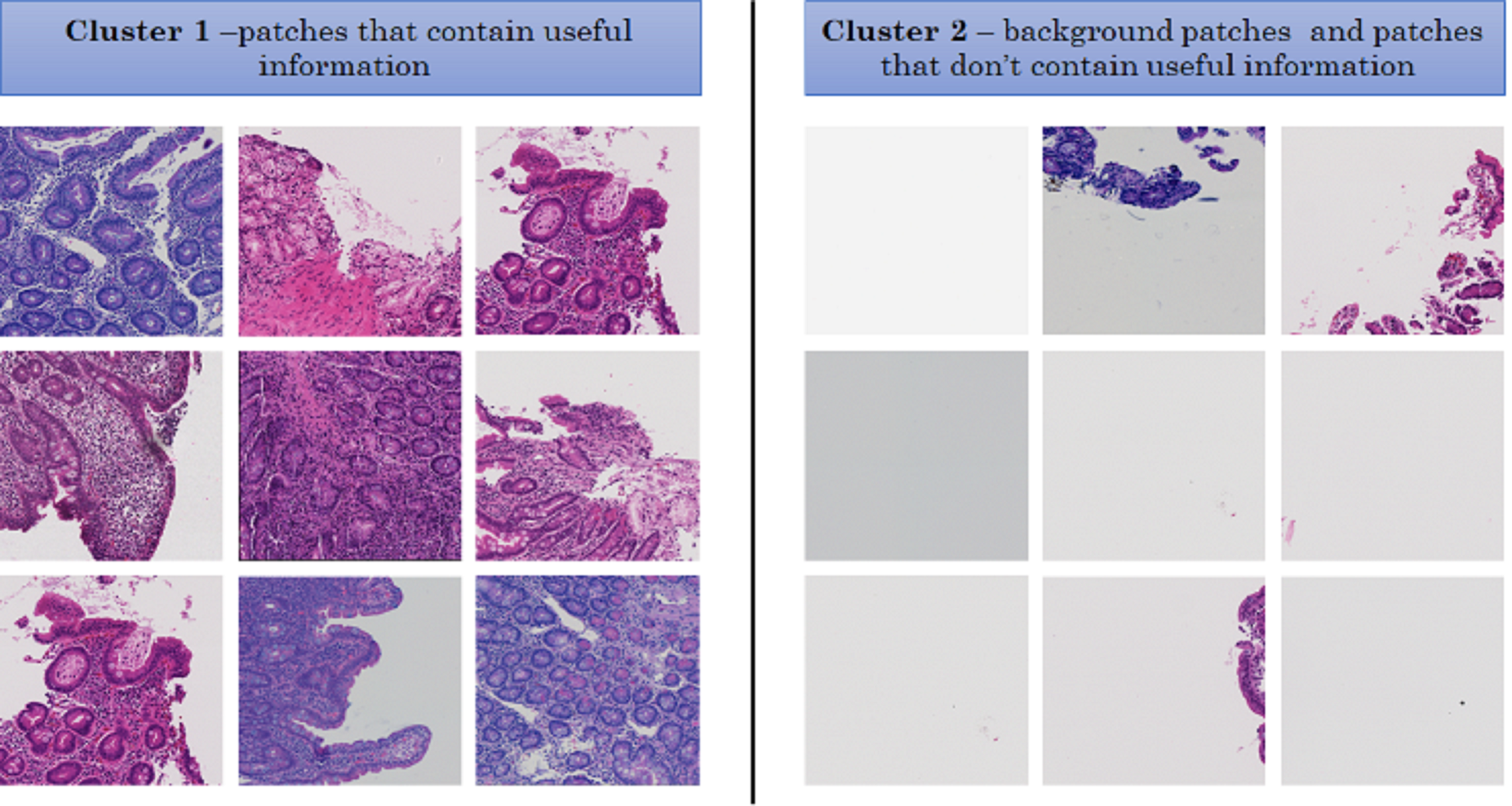}
    \caption{Some samples of clustering results - cluster 1 includes patches with useful information and cluster 2 includes patches without useful information (mostly created from background parts of WSIs)} \label{fig_Clustering}
\end{figure}

\begin{table}[!t]
\caption{The clustering results for all patches into two clusters}\label{tb:clustering}
\centering
\begin{tabular}{|c|c|c|c|}
\hline
                                                                         & Total & Cluster 1   & Cluster 2    \\ \hline
Celiac Disease (CD)                                                      & $16,832$  & $7,742~(46\%)$ & $9,090~(54\%)$  \\ \hline
Normal                                                                   & $15,983$  & $8,953~(56\%)$ & $7,030~(44\%)$  \\ \hline
Environmental Enteropathy~(EE)& $22,625$ & $2,034~(9\%)$   & $20,591~(91\%)$ \\ \hline
Total                                                                    & $55,440$ & $18,729~(34\%)$ & $36,711~(66\%)$ \\ \hline
\end{tabular}
\end{table}

\begin{equation}
    z_m = O_m = a(I * F^{(1)}_{m} + b^{(1)}_m) \quad m = 1, \cdots, m
\end{equation}
where~$b^{(1)}_m$ is the bias, and the number of zeros we want to pad the input with is such that: \text{dim}(I) = \text{dim}(\text{decode}(\text{encode}(I))) Finally, the encoding convolution is equal to:
\begin{equation}
\begin{split}
     O_w = O_h &= (I_w + 2(2k +1) -2) - (2k + 1) + 1 \\&= I_w + (2k + 1) - 1
\end{split}
\end{equation}
The decoding convolution step produces~$n$ feature maps~$z_{m=1,\hdots,n}$. The reconstructed results~$\hat{I}$ is the result of the convolution between the volume of feature maps~$Z=\{z_{i=1}\}^n$ and this convolutional filters volume~$F^{(2)}$~\cite{chen2015page,geng2015high}.
\begin{equation}
    \tilde{I} = a(Z * F^{(2)}_{m} + b^{(2)})
\end{equation}
\begin{equation}\label{eq:a:CNN}
\begin{split}
      O_w = O_h = ( I_w + (2k + 1) - 1 ) -  (2k + 1) + 1 = I_w = I_h  
\end{split}
\end{equation}
Where Equation~\ref{eq:a:CNN} shows the decoding convolution with ~$I$ dimensions. The input's dimensions are equal to the output's dimensions.

 Results of patch clustering has been summarized in Table~\ref{tb:clustering}. Obviously, patches in cluster~$1$, which were deemed useful, are used for the analysis in this paper.

\begin{figure}[!b]
    \centering
    \includegraphics[width=\columnwidth]{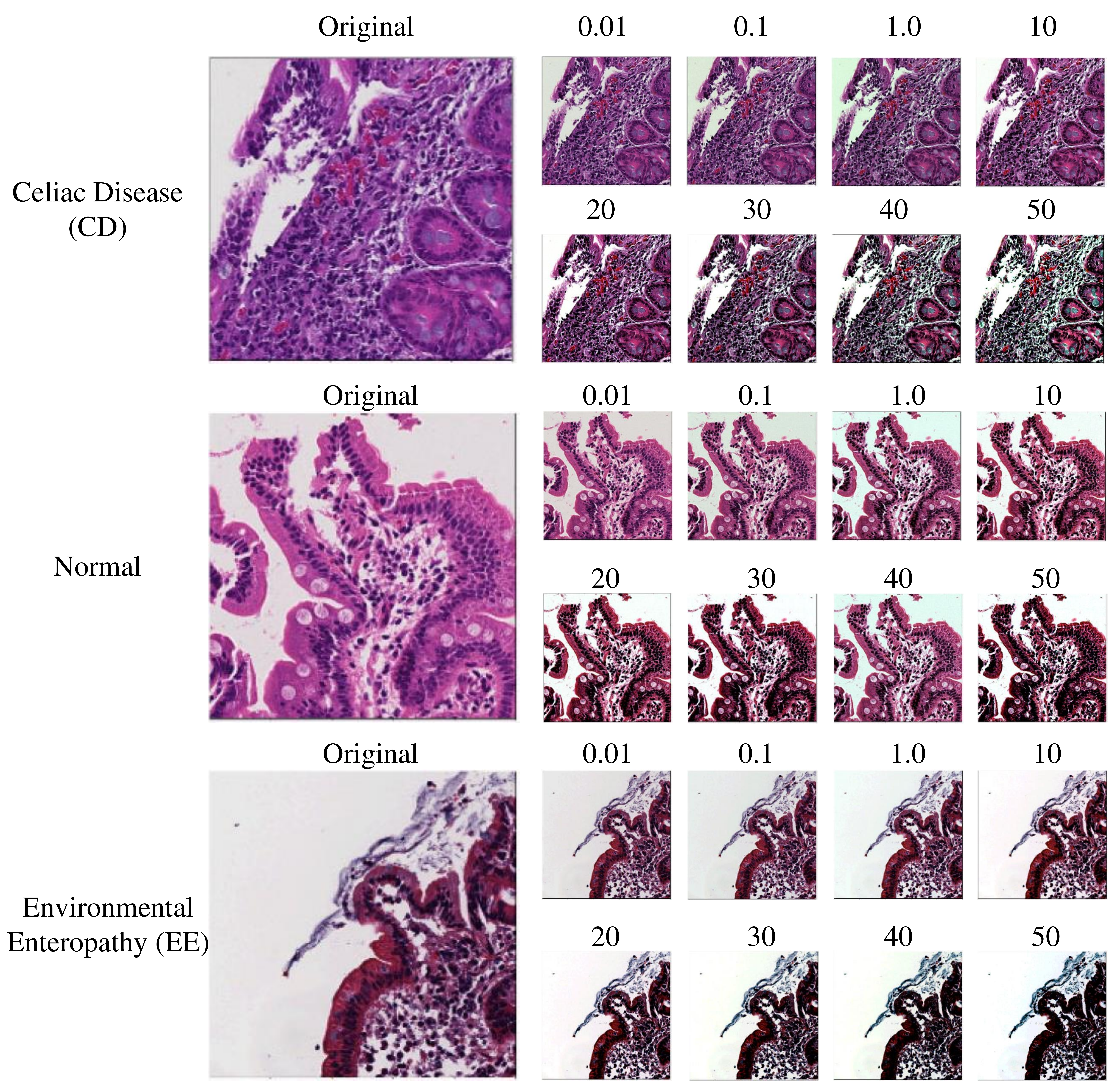}
    \caption{Color Balancing samples for the three classes} \label{fig:CB}
\end{figure}

\subsection{Color Balancing}\label{subsec:CB}
The concept of color balancing for this paper is to convert all images to the same color space to account for variations in H\&E staining. The images can be represented with the illuminant spectral power distribution~$I(\lambda)$, the surface spectral reflectance~$S(\lambda)$, and the sensor spectral sensitivities~$C(\lambda)$~\cite{bianco2017improving,bianco2014error}. Using this notation~\cite{bianco2014error}, the sensor responses at the pixel with coordinates~$(x,y)$ can be thus described as:

\begin{equation}
    p(x,y) = \int_w I(x,y,\lambda) S(x,y,\lambda) C(\lambda) d\lambda
\end{equation}
where~$w$ is the wavelength range of the visible light spectrum, ρ and~$C(\lambda)$ are three-component vectors.

\begin{equation}\label{eq_RGB_IN_OUT}
    \begin{aligned} \left [ \begin{array}{c} R \\ G \\ B \\ \end{array} \right ]_{out} =& \left( \alpha \left [ \begin{array}{c@{\quad}c@{\quad}c} a_{11} & a_{12} & a_{13} \\ a_{21} & a_{22} & a_{23} \\ a_{31} & a_{32} & a_{33}\\ \end{array} \right ]\right. {}\times\left. \left [ \begin{array}{c@{\quad}c@{\quad}c} r_{awb} & 0 & 0 \\ 0 & g_{awb} & 0 \\ 0 & 0 & b_{awb} \\ \end{array} \right ] \left [ \begin{array}{c} R \\ G \\ B \\ \end{array} \right ]_{in} \right)^{\gamma} \end{aligned}
\end{equation}
where\textbf{~$RGB_{in}$}  is raw images from biopsy and~\textbf{~$RGB_{out}$} is results for CNN input. In the following, a more compact version of Equation~\ref{eq_RGB_IN_OUT} is used:
\begin{equation}
    RGB_{out} = (\alpha AI_w . RGB_{in})^\gamma
\end{equation}

where~$\alpha$ is exposure compensation gain, $I_w$ refers the diagonal matrix for the illuminant compensation and~$A$ indicates the color matrix transformation. 

Figure~\ref{fig:CB} shows the results of color balancing for three classes~(Celiac Disease (CD), Normal and Environmental Enteropathy (EE)) with different color balancing percentages between~$0.01$ and~$50$.

\section{Method}\label{sec:Method}
\begin{figure}[t]
    \centering
    \includegraphics[width=\textwidth]{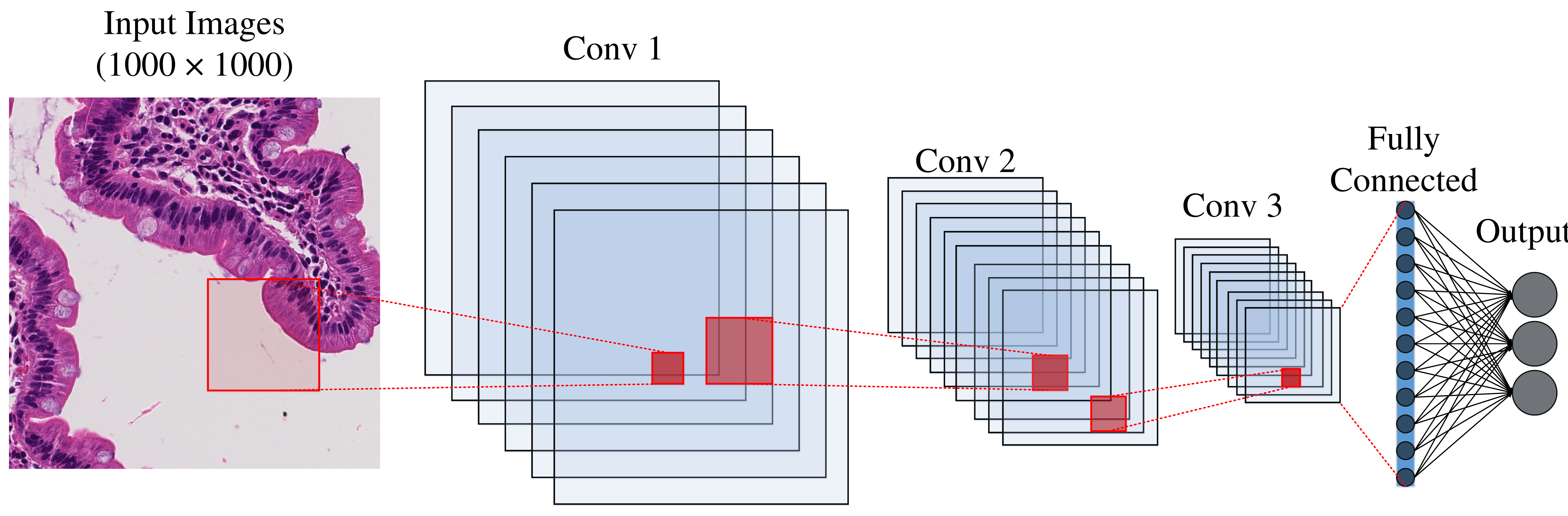}
    \caption{Structure of Convolutional Neural Net using multiple 2D feature detectors and 2D max-pooling} \label{cnn_fig}
\end{figure}

In this section, we describe Convolutional Neural Networks~(CNN) including the convolutional layers, pooling layers, activation functions, and optimizer. Then, we discuss our network architecture for diagnosis of Celiac Disease and Environmental Enteropathy. As shown in figure~\ref{cnn_fig}, the input layers starts with image patches~($1000\times 1000$) and is connected to the convolutional layer~(\textit{Conv~$1$}). Conv~$1$ is connected to the pooling layer~(\textit{MaxPooling}), and then connected to \textit{Conv~$2$}. Finally, the last convolutional layer~(\textit{Conv~$3$}) is flattened and connected to a fully connected perception layer. The output layer contains three nodes which each node represent one class.

\subsection{Convolutional Layer}
CNN is a deep learning architecture that can be employed for hierarchical image classification. Originally, CNNs were built for image processing with an architecture similar to the visual cortex. CNNs have been used effectively for medical image processing. In a basic CNN used for image processing, an image tensor is convolved with a set of kernels of size~$d \times d$. These convolution layers are called feature maps and can be stacked to provide multiple filters on the input. The element~(feature) of input and output matrices can be different~\cite{li2014medical}. The process to compute a single output matrix is defined as follows:

\begin{equation}
    A_{j}=f\left(\sum_{i=1}^{N}I_{i}\ast K_{i,j}+B_{j}\right)
\end{equation}
Each input matrix~$I-i$ is convolved with a corresponding kernel matrix~$K_{i,j}$, and summed with a bias value~$B_j$ at each element. Finally, a non-linear activation function~(See Section~\ref{Sec:Activation}) is applied to each element~\cite{li2014medical}.

In general, during the back propagation step of a CNN, the weights and biases are adjusted to create effective feature detection filters . The filters in the convolution layer are applied across all three 'channels' or $\Sigma$~(size of the color space)~\cite{Heidarysafa2018RMDL}. 

\subsection{Pooling Layer}
To reduce the computational complexity, CNNs utilize the concept of pooling to reduce the size of the output from one layer to the next in the network. Different pooling techniques are used to reduce outputs while preserving important features ~\cite{scherer2010evaluation}. The most common pooling method is max pooling where the maximum element is selected in the pooling window.\\ In order to feed the pooled output from stacked featured maps to the next layer, the maps are flattened into one column. The final layers in a CNN are typically fully connected~\cite{kowsari2018rmdl}.\\

\subsection{Neuron Activation}\label{Sec:Activation}
 The implementation of CNN is a discriminative trained model that uses standard back-propagation algorithm using a sigmoid~(Equation~\ref{sigmoid}), (Rectified Linear Units~(ReLU)~\cite{nair2010rectified}~(Equation~\ref{relu}) as activation function. The output layer for multi-class classification includes a $Softmax$ function~(as shown in Equation~\ref{Softmax}).
 \begin{align}
f(x) &= \frac{1}{1+e^{-x}}\in (0,1)\label{sigmoid}\\
f(x) &= \max(0,x)\label{relu}\\
\sigma(z)_j &= \frac{e^{z_j}}{\sum_{k=1}^K e^{z_k}}\label{Softmax}\\ 
&\forall   ~j \in \{1,\hdots, K\} \nonumber
\end{align}

\subsection{Optimizor}\label{sec:optimizer}
For this CNN architecture, the $Adam$ optimizor~\cite{kingma2014adam} which is a stochastic gradient optimizer that uses only the average of the first two moments of gradient~($v$ and $m$, shown in Equation~\ref{adam}, \ref{adam1}, \ref{adam2}, and \ref{adam3}). It can handle non-stationary of the objective function as in RMSProp, while overcoming the sparse gradient issue limitation of RMSProp~\cite{kingma2014adam}.

\begin{equation}
\theta  \leftarrow \theta - \frac{\alpha}{\sqrt{\hat{v}}+\epsilon} \hat{m}\label{adam}
\end{equation}
\begin{equation}
g_{i,t} =  \nabla_\theta J(\theta_i , x_i,y_i) \label{adam1}
\end{equation}
\begin{equation}
m_t = \beta_1 m_{t-1} + (1-\beta_1)g_{i,t}\label{adam2}
\end{equation}
\begin{equation}
m_t = \beta_2 v_{t-1} + (1-\beta_2)g_{i,t}^2\label{adam3}
\end{equation}
where $m_t$ is the first moment and $v_t$ indicates second moment that both are estimated. $\hat{m_t}=\frac{m_t}{1-\beta_1^t}$ and $\hat{v_t}=\frac{v_t}{1-\beta_2^t}$.\vspace{5.5pt}\\ 
\subsection{Network Architecture}
\begin{figure}[!t]
    \centering
    \includegraphics[width=\textwidth]{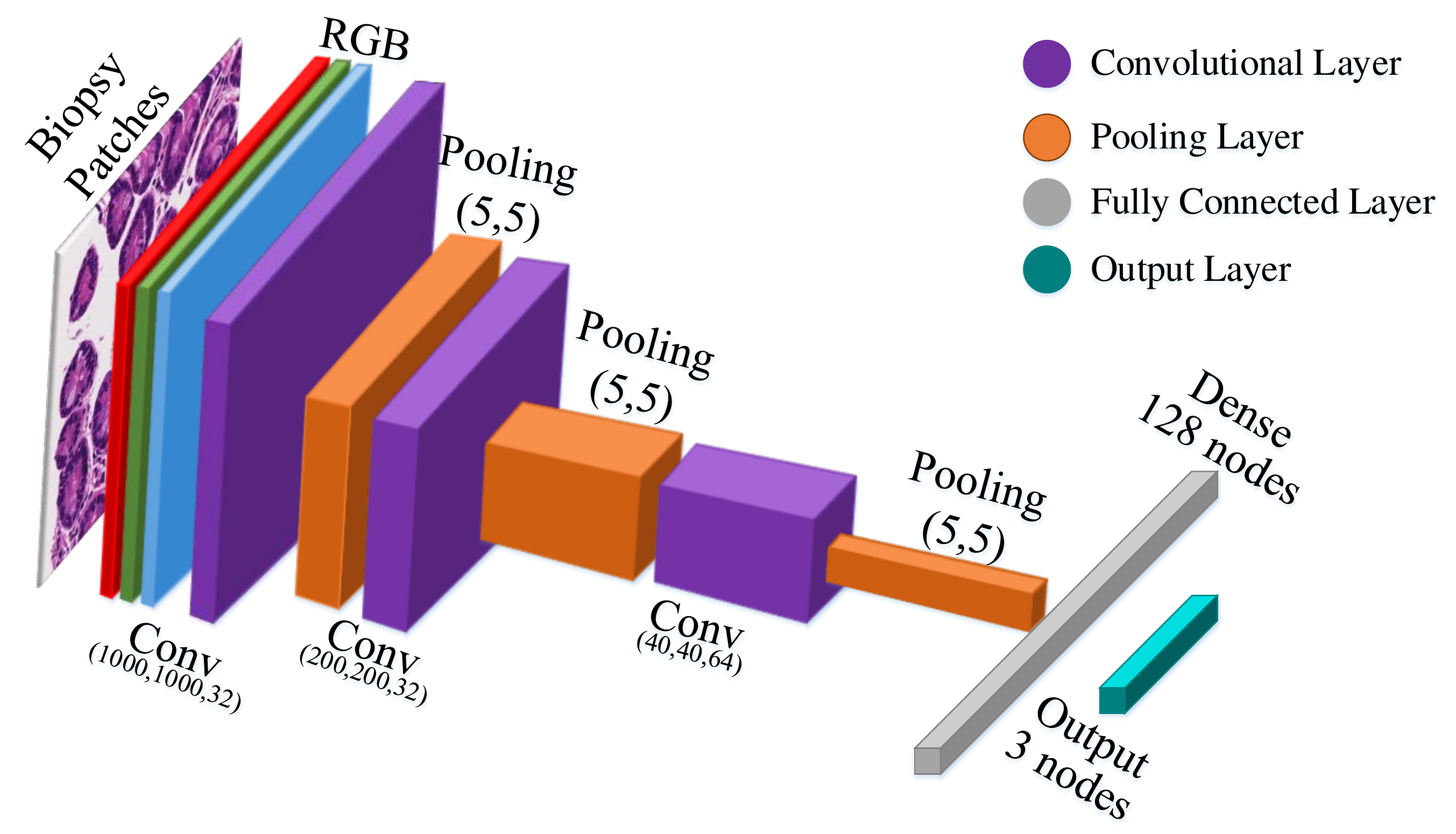}
    \caption{Our Convolutional Neural Networks' Architecture} \label{fig:cnn_Ar}
\end{figure}

As shown in Table~\ref{tb:CNN} and Figure~\ref{fig:cnn_Ar}, our CNN architecture consists of three convolution layer each followed by a pooling layer. This model receives RGB image patches with dimensions of ~$(1000\times 1000)$ as input. The first convolutional layer has~$32$ filters with kernel size of~$(3, 3)$. Then we have Pooling layer with size of~$(5,5)$ which reduce the feature maps from~$(1000\times 1000)$ to~$(200 \times 200)$. The second convolutional layers with~$32$ filters with kernel size of~$(3, 3)$. Then Pooling layer~(MaxPooling~$2D$) with size of~$(5,5)$ reduces the feature maps from~$(200\times 200)$ to~$(40 \times 40)$. The third convolutional layer has~$64$ filters with kernel size of~$(3, 3)$, and final pooling layer~(MaxPooling~$2D$) is scaled down to~$(8 \times 8)$. The feature maps as shown in Table~\ref{tb:CNN} is flatten and connected to fully connected layer with~$128$ nodes. The output layer with three nodes to represent the three classes: ~(Environmental Enteropathy, Celiac Disease, and Normal).

The optimizer used is Adam~(See Section~\ref{sec:optimizer}) with a learning rate of~$0.001$, $\beta_1=0.9$, $\beta_2=0.999$ and the loss considered is sparse categorical crossentropy~\cite{chollet2015keras}. Also for all layers, we use Rectified linear unit~(ReLU) as activation function except output layer which we use~$Softmax$~(See Section~\ref{Sec:Activation}).  

\begin{table}[h]
\centering
\caption{CNN Architecture for Diagnosis of Diseased Duodenal on Biopsy Images}\label{tb:CNN}
\begin{tabular}{|c|c|c|c|}
\hline
   & Layer (type)        & Output Shape     & \begin{tabular}[c]{@{}c@{}}Trainable \\ Parameters\end{tabular} \\ \hline
1  & Convolutional Layer & $(1000, 1000, 32)$ & $869$                                                             \\ \hline
2  & Max Pooling        & $(200, 200, 32)$   & $0$                                                               \\ \hline
3  & Convolutional Layer & $(200, 200, 32)$   & $9,248$                                                            \\ \hline
4  & Max Pooling        & $(40, 40, 32)$     & $0$                                                               \\ \hline
5  & Convolutional Layer & $(40, 40, 64)$     & $18,496$                                                           \\ \hline
6  & Max Pooling        & $(8, 8, 64)$       & $0$                                                               \\ \hline
7  & dense               & $128$              & $524,416$                                                          \\ \hline
8 & Output              & $3$                & $387$                                                             \\ \hline
\end{tabular}
\end{table}

\section{Empirical Results}\label{sec:Empirical_Results}

\subsection{Evaluation Setup}\label{sec:Evaluation}
In the research community, comparable and shareable performance measures to evaluate algorithms are preferable. However, in reality such measures may only exist for a handful of methods. The major problem when evaluating image classification methods is the absence of standard data collection protocols. Even if a common collection method existed, simply choosing different training and test sets can introduce inconsistencies in model performance~\cite{yang1999evaluation}. Another challenge with respect to method evaluation is being able to compare different performance measures used in separate experiments. Performance measures generally evaluate specific aspects of classification task performance, and thus do not always present identical information. In this section, we discuss evaluation metrics and performance measures and highlight ways in which the performance of classifiers can be compared.

Since the underlying mechanics of different evaluation metrics may vary, understanding what exactly each of these metrics represents and what kind of information they are trying to convey is crucial for comparability. Some examples of these metrics include recall, precision, accuracy, F-measure, micro-average, and macro-average. These metrics are based on a~``confusion matrix'' that comprises true positives~(TP), false positives~(FP), false negatives~(FN) and true negatives~(TN)~\cite{lever2016points}. The significance of these four elements may vary based on the classification application. The fraction of correct predictions over all predictions is called accuracy~(Eq. \ref{eq:acc}). The proportion of correctly predicted positives to all positives is called precision,~\textit{i.e.} positive predictive value (Eq. \ref{eq:pres}). 

\begin{equation}
    accuracy=\frac{(TP+TN)}{(TP+FP+FN+TN)}\label{eq:acc}
\end{equation}
\begin{equation}
Precision = \frac{\sum_{l=1}^LTP_l}{\sum_{l=1}^LTP_l+FP_l}\label{eq:pres}
\end{equation}
\begin{equation}
Recall= \frac{\sum_{l=1}^LTP_l}{\sum_{l=1}^LTP_l+FN_l}\label{eq:recall}
\end{equation}
\begin{equation}
F1-Score =  \frac{\sum_{l=1}^L2TP_l}{\sum_{l=1}^L2TP_l+FP_l+FN_l}
\end{equation}

\subsection{Experimental Setup}\label{sec:Evaluation}
The following results were obtained using a combination of central processing units~(CPUs) and graphical processing units~(GPUs). The processing was done on a $Xeon~E5-2640~ (2.6 GHz)$ with $32$ cores and $64 GB$ memory, and the GPU cards were two $Nvidia~Titan~Xp$ and a $Nvidia~Tesla~K20c$. We implemented our approaches in Python using the Compute Unified Device Architecture~(CUDA), which is a parallel computing platform and Application Programming Interface~(API) model created by $Nvidia$. We also used Keras and TensorFlow libraries for creating the neural networks~\cite{abadi2016tensorflow,chollet2015keras}. 

\subsection{Experimental Results}
In this section we show that CNN with color balancing can improve the robustness of  medical image classification. The results for the model trained on $4$ different color balancing values are shown in Table~\ref{tb:1}. The results shown in Table~\ref{tb:2} are also based on the trained model using the same color balancing values. Although in Table~\ref{tb:2}, the test set is based on a different set of color balancing values: ~$0.5, 1.0, 1.5$~and~$2.0$. By testing on a different set of color balancing, these results show that this technique can solve the issue of multiple stain variations during histological image analysis.

As shown in Table~\ref{tb:1}, the f1-score of three classes~(Environmental Enteropathy~(EE), Celiac Disease~(CD), and Normal) are $0.98$, $0.94$, and $0.91$ respectively. In Table~\ref{tb:2}, the f1-score is reduced, but not by a significant amount. The three classes~(Environmental Enteropathy~(EE), Celiac Disease~(CD), and Normal) f1-scores are $0.94$, $0.92$, and $0.87$ respectively. The result is very similar to \textit{MA. Boni et.al}~\cite{Mohammad_al_boni} which achieved 90.59\% of accuracy in their mode, but without using the color balancing technique to allow differently stained images.

\begin{table}[h]
\centering
\caption{F1-score for train on a set with color balancing of 0.001, 0.01, 0.1, and 1.0. Then, we evaluate test set with same color balancing}\label{tb:1}
\begin{tabular}{|c|c|c|c|c|}
\hline
   & precision & recall & f1-score & support \\ \hline
Celiac Disease (CD)  & $0.89$      & $0.99$   & $0.94$     & $22,196$   \\ \hline
Normal  & $0.99$      & $0.83$   & $0.91$     & $22,194$   \\ \hline
\begin{tabular}[c]{@{}c@{}}Environmental  Enteropathy \\ (EE) \end{tabular} & $0.96$      & $1.00$   & $0.98$     & $22,198$   \\ \hline
\end{tabular}
\end{table}

\begin{table}[h]
\centering
\caption{F1-score for train with color balancing of 0.001, 0.01, 0.1, and 1.0 and test with color balancing of 0.5, 1.0, 1.5 and 2.0}\label{tb:2}
\begin{tabular}{|c|c|c|c|c|}
\hline
   & precision & recall & f1-score & support \\ \hline
Celiac Disease (CD)  & $0.90$      & $0.94$   & $0.92$     & $22,196$   \\ \hline
Normal  & $0.96$      & $0.80$   & $0.87$     & $22,194$   \\ \hline
\begin{tabular}[c]{@{}c@{}}Environmental Enteropathy  \\ (EE) \end{tabular} & $0.89$      & $1.00$   & $0.94$     & $22,198$   \\ \hline
\end{tabular}
\end{table}

\begin{figure}[!hbt]
    \centering
    \includegraphics[width=0.7\textwidth]{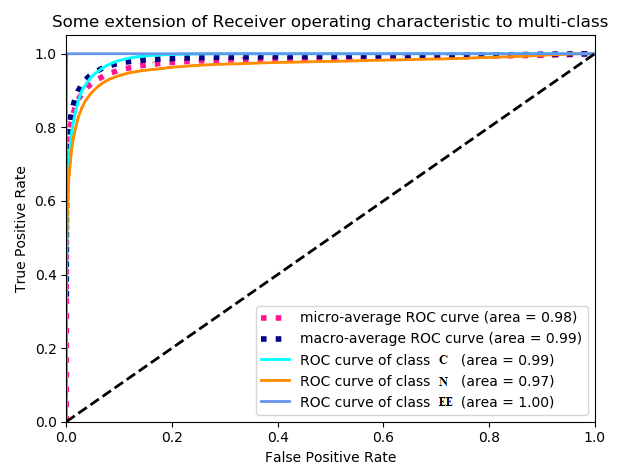}
    \caption{Receiver operating characteristics~(ROC) curves for three classes also the figure shows micro-average and macro-average of our classifier} \label{fig:ROC}
\end{figure}

In Figure~\ref{fig:ROC}, Receiver operating characteristics~(ROC) curves are valuable graphical tools for evaluating classifiers. However, class imbalances (i.e. differences in prior class probabilities) can cause ROC curves to poorly represent the classifier performance. ROC curve plots true positive rate~(TPR) and false positive rate~(FPR). The ROC shows that AUC of Environmental Enteropathy~(EE) is~$1.00$, Celiac Disease~(CD) is 0.99, and Normal is 0.97.

\begin{table}[!hbt]
\caption{Comparison accuracy with different baseline methods}\label{tb:compare}
\begin{tabular}{lccc}
\hline
Method                    & ~~~~\begin{tabular}[c]{@{}c@{}}Solve Color\\  Staining Problem\end{tabular}~~~~~ & ~~~\begin{tabular}[c]{@{}c@{}}Model\\ Architecture\end{tabular}~~~ & ~~~Accuracy~~~       \\ \hline
Shifting and Reflections~\cite{Mohammad_al_boni}  & No                     & CNN                & 85.13\%        \\ 
Gamma~\cite{Mohammad_al_boni}                     & No                     & CNN                & 90.59\%        \\ 
CLAHE~\cite{Mohammad_al_boni}                     & No                     & CNN                & 86.79\%        \\ 
Gamma-CLAHE~\cite{Mohammad_al_boni} & No                     & CNN                & 86.72\%        \\ 

Fine-tuned ALEXNET~\cite{nawaz2018classification}          & \textbf{Yes}           & ALEXNET            & 89.95\%     \\     
Ours                      & \textbf{Yes}            & CNN                & \textbf{93.39\% } \\ \hline
\end{tabular}
\end{table}

As shown in Table~\ref{tb:compare}, our model performs better compared to some other models in terms of accuracy. Among the compared models, only the fine-tuned ALEXNET~\cite{nawaz2018classification} has considered the color staining problem. This model proposes a transfer learning based approach for the classification of stained histology images. They also applied stain normalization before using images for fine tuning the model.

\section{Conclusion}\label{sec:Conclusion}
In this paper, we proposed a data driven model for diagnosis of diseased duodenal architecture on biopsy images using color balancing on convolutional neural networks. Validation results of this model show that it can be utilized by pathologists in diagnostic operations regarding CD and EE. Furthermore, color consistency is an issue in digital histology images and different imaging systems reproduced the colors of a histological slide differently. Our results demonstrate that application of the color balancing technique can attenuate effect of this issue in image classification.  

The methods described here can be improved in multiple ways. Additional training and testing with other color balancing techniques on data sets will continue to identify architectures that work best for these problems. Also, it is possible to extend the model to more than four different color balance percentages to capture more of the complexity in the medical image classification.

\section*{Acknowledgements}

This research was supported by University of Virginia, Engineering in Medicine SEED Grant $(SS~\&~DEB)$, the University of Virginia Translational Health Research Institute of Virginia~($THRIV$) Mentored Career Development Award $(SS)$, and the Bill and Melinda Gates Foundation~~($AA,~OPP1138727$; $SRM$, $OPP1144149$; $PK,~OPP1066118$)

\bibliographystyle{bibtex/spmpsci} 
\bibliography{template.bib}

\end{document}